\documentclass[twocolumn,showpacs,preprintnumbers,amsmath,amssymb,showkeys]{revtex4}

\usepackage{graphicx}
\usepackage{dcolumn}
\usepackage{bm}

\begin{document}

\preprint{}

\title{Tunable invisibility cloaking by using \\ graphene-coated nanowires}

\author{Mahin Naserpour}
 \altaffiliation[Also at ]{Department of Physics, Shiraz University, Iran}
\author{Carlos J. Zapata-Rodr\'{\i}guez}%
 \email{carlos.zapata@uv.es}
\affiliation{%
Department of Optics and Optometry and Vision Science, University of Valencia, \\ 
Dr. Moliner 50, 46100 Burjassot, Spain
}%

\author{Slobodan M. Vukovi\'c}
\altaffiliation[Also at ]{IHTM, University of Belgrade, Serbia}
\author{Milivoj R. Beli\'c}%
\affiliation{%
	Texas A\&M University at Qatar, P.O. Box 23874, Doha, Qatar
}%

\date{\today}

\begin{abstract}
We investigate, both theoretically and numerically, a graphene-coated nano-cylinder illuminated by a plane electromagnetic wave in THz range of frequencies. 
We have derived an analytical formula that enables fast evaluation of the spectral window with a substantial reduction in scattering efficiency for sufficiently thin cylinder. 
This effect leads to tunable resonant invisibility that can be achieved via modification of graphene chemical potential monitored by the gate voltage. 
A multi-frequency cloaking mechanism based on dimer coated nanowires is also discussed.
\end{abstract}

\keywords{Nanowires, Plasmonics, THz Optics}
\maketitle

\section{Introduction}

Recent progress in design and manufacturing of new photonic materials has enabled novel models and multifunctional devices to achieve unprecedented control of light. 
In particular, by using various techniques based on transformation optics, the possibilities have appeared to electromagnetically isolate a space region for certain frequencies or range of frequencies.
It means that an object located inside such a space region will practically stop interacting with illuminating light  \cite{Leonhardt06, Pendry06, Cai07}.
In fact, a remote observer will not be able to detect the presence of the objects that are shielded and protected by optical cloaking. 
A major drawback in practical applications of these techniques is due to the properties of materials needed, either artificial or natural, that exhibit extreme permittivity or permeability. 
However, coatings can be designed such as metal-dielectric multilayer or metasurfaces to drastically reduce the scattered signal from subwavelength particles and make the assembly nearly undetectable \cite{Filonov12, Chen12, Monti15, Soric15}.
Of course, the light interacts with the nanoparticles, but the destructive interference of different elements that constitute the complex nanostructure practically cancels out the scattered radiation.

Graphene represents a groundbreaking and exciting material that combines suitable characteristics for the use in optoelectronic devices in far-infrared and THz region of frequencies due to the intrinsic ability to mold the surface current with low-loss rates \cite{Zhu10,Bonaccorso10}. 
High conductivity within atomically thin layer of graphene, and significant tuning ability via the applied bias voltage, enables the cancellation of scattering effects. 
Drastically reduced overall visibility of the scattering object that was conducted via graphene monolayer wrapping the cylinder, for the first time, was extended to the far-infrared range of frequencies by Chen and Alu \cite{Chen11c}. 
In that case, the local polarizability of a nanowire with moderately thick diameter and the graphene coating sheet with the opposite signs can be mutually cancelled under suitable layout. 
Different schemes have been proposed with plasmonic compounds, using a dipole moment of opposite phase to attain a scattering cancellation \cite{Alu05,Edwards09,Tricarico09}. 
If the incident radiation is $p$-polarized, however, the spectral response of the scattering cross section seems to be totally dissimilar \cite{Riso15}. 
Then, a set formed by a peak, associated with the scattering via a narrow Mie mode, and a valley offers the prospect to obtain invisibility with unprecedented control over the wire resonance spectral location by changing the chemical potential.

In the present paper we extend the previous concept by employing a $p$-polarized incident plane wave that illuminates a graphene coated dielectric nanowire in order to get near the fundamental localized surface plasmon polariton. 
Subsequently we optimize the scattering cancellation of the nanocavity. 
The proposed theoretical model is critically discussed with a conclusion that a characteristic lineshape appears comparable to the shape of Fano resonance \cite{Lukyanchuk10}. 
The check of validity and estimation of our results is performed by using the full-wave Lorenz-Mie method \cite{Bohren98}.

\section{Formulation of the problem. Analytical considerations}

\begin{figure}[t]
	\centering
	\includegraphics[width=0.75\linewidth]{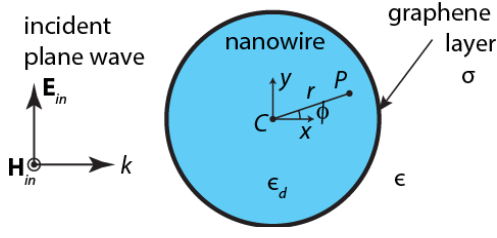}
	\caption{Illustration of the atomically thin graphene layer of conductivity $\sigma$ coating a nano-sized scattering cylinder of permittivity $\epsilon_d$, where the environment medium has a permittivity $\epsilon$.
		Here we consider an incident TE$^z$-polarized plane wave.}
	\label{fig01}
\end{figure}

Let us consider a cylindrical dielectric nanowire of radius $R$ and relative permittivity $\epsilon_d$ over-coated by an atomically thin graphene shell, and oriented along $z$ axis, as illustrated in Fig.~\ref{fig01}. 
The nanowire is assumed to be placed in a lossless environment medium of the relative permittivity $\epsilon$ and illuminated by a plane electromagnetic wave that propagates along the $x$ axis with the electric field vector $\mathbf{E}_{in}$ in the $xy$ plane. 
The graphene conductivity $\sigma$ within the coating shell is described in the local random phase approximation by using Kubo formula \cite{Nikitin11} as a sum of intraband $\sigma_\mathrm{intra}$ and interband $\sigma_\mathrm{inter}$ contribution, where
\begin{equation}
\sigma_\mathrm{intra} = \frac{2 i e^2 k_B T}{\pi \hbar^2 (\omega + i \Gamma)} \ln \left[ 2 \cosh \left( \frac{\mu}{2 k_B T} \right) \right] ,
\label{eq12}
\end{equation}
and 
\begin{widetext}
\begin{equation}
\sigma_\mathrm{inter} = \frac{e^2}{4 \hbar} \left[ \frac{1}{2} + \frac{1}{\pi} \arctan \left( \frac{\hbar \omega - 2 \mu}{2 k_B T} \right)
- \frac{i}{2 \pi} \ln \frac{\left( \hbar \omega + 2 \mu \right)^2}{\left( \hbar \omega - 2 \mu \right)^2 + \left( 2 k_B T \right)^2} \right] .
\end{equation}
\end{widetext}
Here, $-e$ is the charge of an electron, $\hbar$ is the reduced Planck's constant, $k_B$ is Boltzmann's constant, $T$ is the temperature, $\Gamma$ is the charge carriers scattering rate, $\mu$ is the chemical potential and $\hbar \omega$ is the photon energy.
Also, the time harmonic dependence of the field is adopted to be of the form $\exp \left( -i \omega t \right)$.

To estimate analytically the scattering efficiency of the coated nanowire, we followed the Lorenz-Mie scattering method given in detail for instance in Refs.~\cite{Shah70,Bussey75,Chen11c,Riso15,Velichko16}.
As mentioned above, we assume that the nanotube is illuminated by a TE$^z$-polarized plane wave propagating along the $x$ axis, as illustrated in Fig.~\ref{fig01}.
The electromagnetic field of the incident plane wave may be set as 
\begin{equation}
\mathbf{H}_{in} = \hat{z} H_0 \exp \left( i k x \right) = \hat{z} H_0 \sum_{n = - \infty}^{+\infty} i^n J_n \left(k r \right) \exp \left( i n \phi \right) ,
\label{eq03}
\end{equation}
and $\mathbf{E}_{in} = i {\nabla \times \mathbf{H}_{in}}/{\omega \epsilon \epsilon_0}$, which reads as
\begin{equation}
\mathbf{E}_{in} = - E_0 \sum_{n=-\infty}^{+\infty} i^n \left[ \hat{r} n \frac{J_n \left( kr \right)}{k r} 
+ \hat{\phi} i J'_n \left( kr \right) \right] \exp \left( i n \phi \right) ,
\end{equation}
where $r$ and $\phi$ are the radial and azimuthal cylindrical coordinates, respectively, $H_0$ is a constant amplitude, $E_0 = {H_0 k} / { \omega \epsilon \epsilon_0}$, $J_n(\cdot)$ is the Bessel function of the first kind and order $n$, $k_0 = \omega/c$ is the wavenumber in the vacuum, and $k = k_0 \sqrt{\epsilon}$.
Here the prime appearing in $J'_n \left( \alpha \right)$ denotes derivative with respect to the variable $\alpha$.
In Eq.~(\ref{eq03}) we used the Jacobi-Anger expansion of a plane wave in a series of cylindrical waves.
The scattered electromagnetic field in the environment medium, $r > R$, may be set as \cite{Bohren98}
\begin{equation}
\mathbf{H}_{sca} = \hat{z} H_0 \sum_{n = - \infty}^{+\infty} a_n i^n H_n^{(1)} \left(k r \right) \exp \left( i n \phi \right) ,
\label{eq06}
\end{equation}
and
\begin{eqnarray}
\mathbf{E}_{sca} = - E_0 \sum_{n=-\infty}^{+\infty} a_n i^n \left[ \hat{r} n \frac{H_n^{(1)} \left( k r \right)}{k r}   
+ \hat{\phi} i {H_n^{(1)}}' \left( kr \right) \right] \nonumber \\
\times \exp \left( i n \phi \right),
\end{eqnarray}
where $H_n^{(1)}(\cdot) = J_n(\cdot) + i Y_n(\cdot)$ is the Hankel function of the first kind and order $n$, and the coefficients $a_n$ must be determined.
The total magnetic field in the environment medium is simply $\mathbf{H}_{tot} = \mathbf{H}_{in} + \mathbf{H}_{sca}$.
Finally, the electromagnetic field in the dielectric core of the coated cylinder ($r < R$) is expressed as
\begin{equation}
\mathbf{H}_{d} = \hat{z} H_0 \sum_{n = - \infty}^{+\infty} b_n i^n J_n \left(k_d r \right) \exp \left( i n \phi \right) ,
\end{equation}
and
\begin{eqnarray}
\mathbf{E}_{d} = - \frac{k_d H_0}{ \omega \epsilon_d \epsilon_0} \sum_{n=-\infty}^{+\infty} b_n i^n \left[ \hat{r} n \frac{J_n \left( k_d r \right)}{k_d r}  
+ \hat{\phi} i J'_n \left( k_d r \right) \right] \nonumber \\
\times \exp \left( i n \phi \right),
\label{eq16}
\end{eqnarray}
where the wavenumber $k_d = k_0 \sqrt{\epsilon_d}$.

The Lorenz-Mie scattering coefficients $a_n$ and $d_n$, are determined by means of the proper boundary conditions.
Due to the existence of graphene surface conductivity, the boundary conditions at $r = R$ are given by \cite{Gao14} in order to ensure the continuity of tangential components of the electromagnetic field at the graphene shell:
\begin{eqnarray}
\hat{\phi} \cdot \mathbf{E}_{d} &=& \hat{\phi} \cdot \left(\mathbf{E}_{sca} + \mathbf{E}_{in} \right) , \\
\hat{z} \cdot \mathbf{H}_{d}    &=& \hat{z} \cdot \left(\mathbf{H}_{sca} + \mathbf{H}_{in} \right) + \hat{\phi} \cdot \mathbf{E}_{d} \sigma ,
\end{eqnarray} 
Since the electric field is assumed to be in the $xy$ plane, the $\phi$-component of the effective conductivity has to be taken into account, only.
These two equations can be written as
\begin{eqnarray}
\frac{k_d}{ \epsilon_d} b_n J'_n \left( k_d R \right) 
&=& 
\frac{k}{ \epsilon} \left[ J'_n \left( k R \right) + a_n {H_n^{(1)}}' \left( k R \right) \right] , \\
b_n J_n \left(k_d R \right)
&=& 
J_n \left(k R \right) + a_n H_n^{(1)} \left(k R \right) \nonumber \\
&-& \sigma \frac{i k_d}{ \omega \epsilon_d \epsilon_0} b_n J'_n \left( k_d R \right) .
\end{eqnarray}
Here, the coefficients with different index $n$ can be treated separately due to the linear independence of the multipolar components of the wave field.
Finally, we obtain the analytical expressions for the scattering coefficients \cite{Riso15}:
\begin{widetext}
\begin{eqnarray}
a_n = \frac{k_d J'_n (k_d R) \left[ \epsilon J_n (k R) - i \frac{k \sigma}{\omega \epsilon_0} J'_n (k R) \right] - k \epsilon_d J_n (k_d R) J'_n (k R)}
{k {H_n^{(1)}}' (k R) \left[ \epsilon_d J_n (k_d R) + i \frac{k_d \sigma}{\omega \epsilon_0} J'_n (k_d R) \right] - k_d \epsilon H_n^{(1)} (k R) J'_n (k_d R)},
\label{eq08} \\
b_n = \frac{k \epsilon_d \left[ {H_n^{(1)}}' (k R) J_n (k R) - H_n^{(1)} (k R) J'_n (k R) \right]}
{k {H_n^{(1)}}' (k R) \left[ \epsilon_d J_n (k_d R) + i \frac{k_d \sigma}{\omega \epsilon_0 } J'_n (k_d R) \right] - k_d \epsilon H_n^{(1)} (k R) J'_n (k_d R)} .
\end{eqnarray}
\end{widetext}

\section{Scattering efficiency}

\begin{figure}[t]
	\centering
	\includegraphics[width=0.75\linewidth]{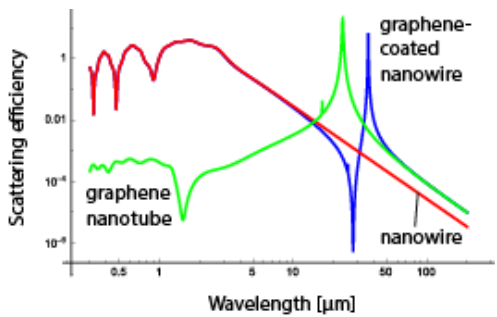}
	\caption{Scattering efficiency, $Q_{sca}$, of a bare cylinder of radius $R = 0.5\ \mu\mathrm{m}$ and permittivity $\epsilon_d = 3.9$ immersed in air, a graphene nanotube of chemical potential $\mu = 0.5\ \mathrm{eV}$ surrounded by air, and the graphene-coated nanocylinder.
		In all calculations we set $T = 300\ \mathrm{K}$ and $\Gamma = 0.1\ \mathrm{meV}$.}
	\label{fig02}
\end{figure}

The scattering coefficients $a_n$, given by Eq.~(\ref{eq08}), enable the fast estimation of the scattering efficiency \cite{Bohren98}:
\begin{equation}
Q_{sca} = \frac{2}{k R} \sum_{n = - \infty}^{+\infty} |a_n|^2 .
\label{eq07}
\end{equation}
Typically, the terms with low order $n$ effectively contribute to the summation when the size of the scattering object is small enough.
The invisibility condition is achieved when the scattering coefficients $a_n$ simultaneously tend to zero.
In contrast, resonances can be attributed to the pole(s) of scattering coefficient(s).
Now, we would like to compare the scattering efficiency of: (i) a bare dielectric cylinder, (ii) a graphene nanotube, and (iii) a uniform-graphene coated dielectric nanocylinder. 
We assume the dielectric core made of SiO$_2$ with the relative permittivity $\epsilon_d = 3.9$ and radius $R = 0.5\ \mu\mathrm{m}$; the graphene monolayer with the chemical potential $\mu = 0.5\ \mathrm{eV}$; ambient temperature $T = 300\ \mathrm{K}$ and carrier scattering rate $\Gamma = 0.1\ \mathrm{meV}$.
The last two parameters are kept fixed in all numerical calculations that follow.
The scattering efficiency spectrum $Q_{sca}(\lambda)$, where $\lambda$ is the wavelength of the TE$^z$-polarized illuminating radiation, for each of the three objects is presented in Fig.~\ref{fig02}.
In the range of long wavelengths, the curve pattern is formed by a peak resonance at $\lambda = 36.4\ \mu\mathrm{m}$ and a minimum of scattering efficiency at a wavelength of $28.1\ \mu\mathrm{m}$.
The resonance wavelength and the invisibility wavelength can be found in the vicinities of the resonance peak for the free-standing graphene nanotube, found at $\lambda = 23.4\ \mu\mathrm{m}$.
Without the presence of the graphene coating, the dielectric cylinder cannot exhibit any significant feature of the scattering efficiency in such spectral range.
One might presume that the set of peak and valley patterned in the scattering efficiency follows the characteristic lineshape of the Fano resonance, where the emission of electromagnetic waves by the coated nanowire creates the interference between the nonresonant scattering from the dielectric core and scattering by narrow Mie modes in the graphene nanotube \cite{Arruda15}.
The graphene-coated nanowire exhibits a scattering efficiency dominated by the graphene nanotube (surrounded by air) at long wavelengths, where the graphene coating isolates the central dielectric cylinder.
On the other hand, the bare (uncoated) nanowire dominates the scattering properties for shorter wavelengths, due to its stronger response far from the Rayleigh scattering regime.
Note that in Fig.~\ref{fig02}, additional minima of the scattering efficiency can be found in the visible and near-IR due to the contribution of high-order Fano resonances, as described in Refs.~\cite{Mirzaei15,Rybin15}.

It is necessary to point out that evaluation of the electromagnetic field components, given by (\ref{eq06}--\ref{eq16}), as well as the scattering efficiency (\ref{eq07}), can be alternatively achieved by using a matrix method to derive the scattering coefficients $a_n$ and $b_n$ \cite{Diaz16d,Diaz16c}.
In this respect, the finite width of the graphene layer ($t_g$) has to be introduced, that leads to the corresponding graphene permittivity $\epsilon_g = 1 + i {\sigma}/{\epsilon_0 \omega t_g}$. 
In particular, $t_g = 0.5\ \mathrm{nm}$ has been employed, where the calculated scattering efficiency for multilayered nanotubes (not shown here) is in excellent agreement with the results depicted in Fig.~\ref{fig02}. 

\begin{figure}[t]
	\centering
	\includegraphics[width=0.75\linewidth]{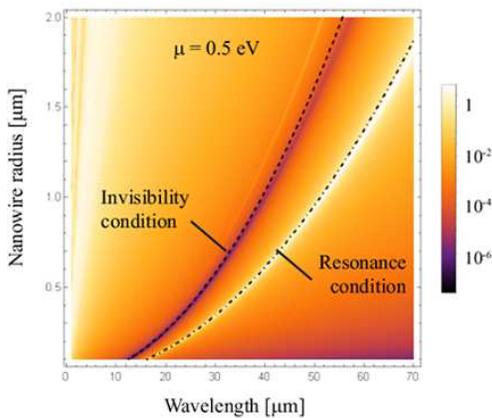}
	\caption{Scattering efficiency spectra, $Q_{sca}$, for a uniform-graphene coated cylinder illuminated by a TE$^z$-polarized plane wave, when the radius $R$ varies, maintaining fixed $\epsilon_d = 3.9$ and $\mu = 0.5\ \mathrm{eV}$.
		We included analytical equations (\ref{eq14}) and (\ref{eq15}) indicating the resonance condition and the invisibility condition, respectively.}
	\label{fig03}
\end{figure}

Consequently, the presence of a peak resonance for a graphene nanotube induces the Fano-like resonance in the complex graphene-dielectric nanowire.
Therefore, tuning the resonance peak of the graphene nanotube will enable to shift the invisibility wavelength on demand.
This procedure can be carried out by simply modifying the radius of the graphene nanotube \cite{Cuevas16}.
Scattering efficiency for a graphene-coated dielectric nanowire as a function of radius $R$ is presented in Fig.~\ref{fig03}. 
It is evident that peak-valley lineshape is preserved, but shifted towards the longer wavelengths when the radius of the cylinder $R$ increases.

On the other side, in order to obtain an analytical expression for the fast evaluation of the invisibility window we use the following approximate expression for $\sigma_\mathrm{intra}$ that dominates the contribution to graphene conductivity for moderate and low frequencies ($\hbar \omega < \mu$), and large doping ($\mu \gg k_B T$):
\begin{equation}
\sigma_\mathrm{intra} = \frac{i e^2 \mu}{\pi \hbar^2 (\omega + i \Gamma)} \ (\approx \sigma) ,
\label{eq1n}
\end{equation}
In fact, this expression represents generalization of the well known Drude model for application to graphene. 
For example, that model is particularly accurate for the imaginary part of graphene conductivity for $\lambda > 10\ \mu\mathrm{m}$, if all the other relevant parameters stay the same as in Fig.~\ref{fig02} \& \ref{fig03}.
In addition, the requirement $\omega \gg \Gamma$ (or $\lambda \ll 12.4\ \mathrm{mm}$) gives the upper limit on the wavelength. 
Therefore, for the parameters used in the present paper, we may conclude that the expression:
\begin{equation}
\mathrm{Im} (\sigma) \approx \frac{e^2 \mu}{\pi \hbar^2 \omega} , 
\end{equation}
can be used with the high accuracy.
On the other hand, analytical approximations for the resonance wavelength and the invisibility wavelength can be deduced by means of the electrostatics approximation \cite{Bohren98,Cuevas16}.
By comparing the squared modulus of the scattering coefficients $a_n$ for nanosized graphene-coated cylinders, one can observe that the first order dominates over the rest of orders in the spectral range of interest.
For $|a_1|^2$ and $R = 0.5\ \mu\mathrm{m}$, we observe a maximum at $36.4\ \mu\mathrm{m}$ and a minimum at $28.1\ \mu\mathrm{m}$, two features which are replicated in the scattering efficiency.
In the latter, in fact an additional peak is also observed at $25.7\ \mu\mathrm{m}$ corresponding to a resonance peak of $|a_2|^2$.
Further peaks and minima are difficult to be registered.
For that reason, the scattering orders $n = \pm 1$ can be taken into account, only.
Within the electrostatics approximation, we further use a Taylor expansion in the limit $R \to 0$, retaining only the term of the lowest order, such as $J_1 (x) \approx x/2$ and $H_1^{(1)} (x) \approx -2 i / \pi x$.
Then,
\begin{equation}
a_1 = \frac{i \pi}{4} \left( k R \right)^2 \frac{\epsilon_d - \epsilon + i \sigma / \epsilon_0 \omega R}{\epsilon_d + \epsilon + i \sigma / \epsilon_0 \omega R} .
\label{eq10}
\end{equation}
In other words, the complex scattering object under consideration behaves like a solid cylinder of dielectric constant $\epsilon_d + i \sigma / \epsilon_0 \omega R$, that assumes a bulk conductivity $\sigma / R$.

When $\omega \gg \Gamma$, it follows from Eq.~(\ref{eq10}) that the resonance appears at the frequency:
\begin{equation}
\omega_\mathrm{res} = \sqrt{\frac{e^2 \mu}{\pi \hbar^2 \epsilon_0 R (\epsilon_d + \epsilon)}} ,
\label{eq14}
\end{equation}
Assuming that $\epsilon_d$ and $\epsilon$ are positive, a resonance peak will occur since $\mathrm{Im} (\sigma) > 0$ in the spectral range of interest.
Note that the peak resonance is associated to a localized surface plasmon in the graphene-coated cylinder, and its dispersion relation can be interpreted such that this bound mode cyclically propagates along the cylinder perimeter exactly one effective wavelength corresponding to the flat graphene sheet \cite{Cuevas16}.
More interesting in our study, a drastic reduction of the scattering efficiency will be experienced under the constrain $\epsilon_d - \epsilon - \mathrm{Im} (\sigma) / \epsilon_0 \omega R = 0$.
The frequency for invisibility will surge at
\begin{equation}
\omega_\mathrm{inv} = \sqrt{\frac{e^2 \mu}{\pi \hbar^2 \epsilon_0 R (\epsilon_d - \epsilon)}} ,
\label{eq15}
\end{equation}
As a consequence, the invisibility window depends on the parameter $\mu/ R \Delta \epsilon$, where $\Delta \epsilon$ stands for the difference of dielectric constants between the core and the environment media.
In order to meet either the invisibility or the scattering resonance conditions, for given monochromatic field, the nanowire radius $R$ has to be adjusted. 
This is illustrated in Fig.~\ref{fig03}, where the analytical expressions (\ref{eq14}) and (\ref{eq15}) have been included for resonance and invisibility, respectively. 
Excellent agreement is evident, especially for low radius nanowires. 
Finally, we would like to emphasise that the scattering efficiency can be accurately calculated as $Q_{sca} = 4 |a_1|^2 / k R$ for sufficiently thin graphene-coated nanowires, that is when the dipolar term given within the electrostatics approximation in Eq.~(\ref{eq10}) dominates.
Therefore, the spectral dependence of the scattering efficiency can be written as 
\begin{equation}
Q_\mathrm{sca} \propto \omega^3 \frac{\left( \omega^2 - \omega_\mathrm{inv}^2 \right)^2 + \left( \omega \Gamma \right)^2}
{\left( \omega^2 - \omega_\mathrm{res}^2 \right)^2 + \left( \omega \Gamma \right)^2} .
\end{equation}
Interestingly, when the regions of invisibility and resonance start to overlap, i.e. when $\omega_\mathrm{inv} - \omega_\mathrm{res} \ll \omega$ in the spectral window of interest, $Q_\mathrm{sca}$ approaches the well known Fano resonance lineshape \cite{Lukyanchuk10}:
\begin{equation}
Q_\mathrm{sca} \propto \frac{\left( F \Gamma / 2 + \omega - \omega_\mathrm{res} \right)^2}
{\left( \omega - \omega_\mathrm{res} \right)^2 + \left( \Gamma / 2 \right)^2} .
\end{equation}
Here, the so called Fano parameter $F = \left( \omega_\mathrm{res} - \omega_\mathrm{inv} \right) / \left( \Gamma / 2 \right)$. 
The validity of such an approximation is limited to the case $\epsilon_d \gg \epsilon$, as discussed below. 
In the case of anisotropic plasmonic nanotubes such behaviour has been highlighted in Ref.~\cite{Diaz16b}.

\begin{figure}[t]
	\centering
	\includegraphics[width=0.9\linewidth]{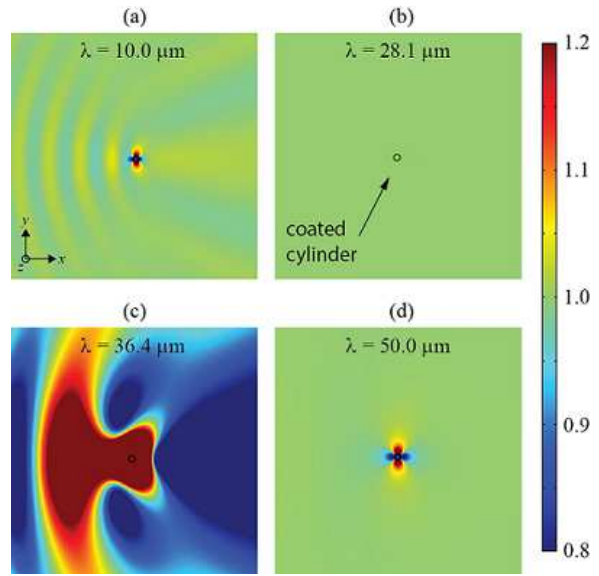}
	\caption{Modulus of the normalized total electric field, $|\mathbf{E}/E_0|$, for a graphene-coated cylinder of radius $R = 0.5\ \mu\mathrm{m}$, keeping $\epsilon_d = 3.9$ and $\mu = 0.5\ \mathrm{eV}$,  illuminated by a TE$^z$-polarized plane wave of wavelength (a) $\lambda = 10.0\ \mu\mathrm{m}$, (b) $\lambda = 28.1\ \mu\mathrm{m}$ coinciding with the invisibility window, (c) $\lambda = 36.4\ \mu\mathrm{m}$ satisfying the resonance condition, and (d) $\lambda = 50.0\ \mu\mathrm{m}$.}
	\label{fig04}
\end{figure}

In Fig.~\ref{fig04} we show the total electric field inside a coated cylinder, $\mathbf{E}_{d}$, and the field in the enviroment medium, $\mathbf{E}_{in} + \mathbf{E}_{sca}$, normalized to the electric field of the incident plane wave, $E_0$, at different wavelengths.
Out of the spectral band where the invisibility and resonance exist, as illustrated for (a) $\lambda = 10\ \mu\mathrm{m}$, and (d) $\lambda = 50\ \mu\mathrm{m}$, a moderately intense signal that corresponds to a dipolar scattered field is evident. 
In this case, Rayleigh scattering is governing, which is characterized by a dominant coefficient $a_1 \propto \lambda^{-2}$.
For $\omega \ll \omega_\mathrm{res}$ the graphene layer acts as a perfect electric conductor (PEC), and the graphene-coated dielectric cylinder behaves exactly like a bare PEC nanowire.
In that case, $a_1 \approx i \pi \left( k R \right)^2 / 4$.
In contrast, for longer wavelengths graphene conductivity can be neglected and the complex nanowire behaves like an uncoated cylinder. 
In that case, $a_1 \approx {i \pi} \left( k R \right)^2 {(\epsilon_d - \epsilon)} / 4 {(\epsilon_d + \epsilon)}$.
The invisibility window is manifested at the wavelength $\lambda = 28.1\ \mu\mathrm{m}$, where a drastic reduction of the scattering wave field is apparent.
Here the first-order scattering coefficient can straightforwardly approximated by $a_1 \approx - \pi \omega_\mathrm{inv} \Gamma \Delta \epsilon R^2 / 8 c^2$, and therefore the scattering efficiency increases with the cylinder radius as $Q_\mathrm{sca} \propto R^3$ at $\omega_\mathrm{inv}$.
Note that the electric field is conserved inside and outside the coated nanowire.
On the contrary, the resonant condition will be satisfied nearby at $\lambda = 36.4\ \mu\mathrm{m}$, where an enhanced scattered field is emitted by the coated cylinder.

\section{Tuning invisibility}

\begin{figure}[t]
	\centering
	\includegraphics[width=0.75\linewidth]{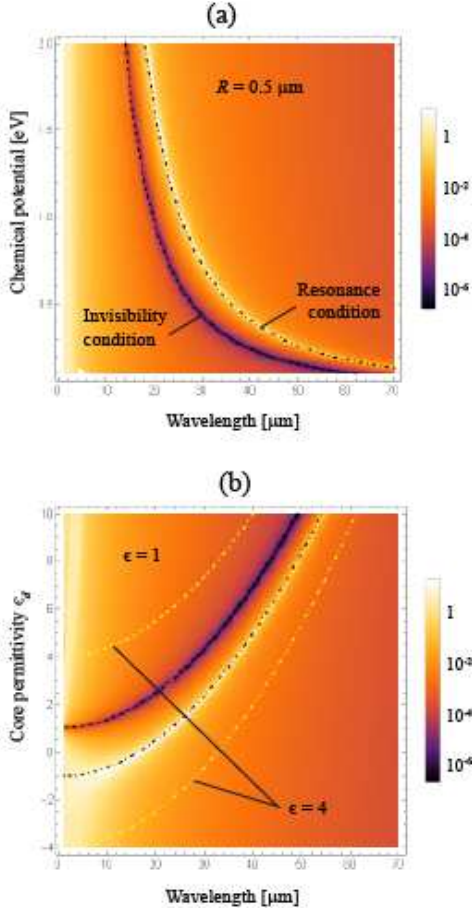}
	\caption{Contour plot of $Q_{sca}$ for graphene coated cylinders illuminated by a TE$^z$-polarized plane wave, when
		(a) the chemical potential $\mu$ is varied, keeping $\epsilon_d = 3.9$ and $R = 0.5\ \mu\mathrm{m}$, and 
		(b) the permittivity $\epsilon_d$ of the dielectric core varies and the environment medium has a dielectric constant $\epsilon = 1$, maintaining fixed $R = 0.5\ \mu\mathrm{m}$ and $\mu = 0.5\ \mathrm{eV}$.
		We also represent the curve of resonance given by Eq.~(\ref{eq14}) in dot-dashed line, and the invisibility condition of Eq.~(\ref{eq15}) drawn in dashed line, when the external medium has a permittivity $\epsilon = 1$ (in black) and, for comparison, when $\epsilon = 4$ (in yellow).}
	\label{fig05}
\end{figure}

A possibility of tuning the invisibility window, as well as the window of the resonance peak, by modifying the chemical potential of the graphene coating that can be monitored by gate voltage seems to be very important for applications. 
The spectrum of $Q_{sca}$ for a varying chemical potential $\mu$ is shown in Fig.~\ref{fig05}(a), assuming a fixed core of radius $R = 0.5\ \mu\mathrm{m}$ and permittivity $\epsilon_d = 3.9$.
We observe that the peak-to-valley sinuosity governing the scattering spectrum is blueshifted for an increasing value of $\mu$.
As long as the chemical potential is enlarged, the graphene conductivity also grows producing a shielding effect in the nanowire, which can be compensated by shifting the working wavelength to the near infrared.
For example, the invisibility spectral band is  $42.25\ \mu\mathrm{m}$ wide when changing the chemical potential from $0.1\ \mathrm{eV}$ for $\lambda_\mathrm{inv} = 62.25\ \mu$m to $1.0\ \mathrm{eV}$ for $\lambda_\mathrm{inv} = 20.00\ \mu\mathrm{m}$.  
Notice, however, that the spectral width of the invisibility window, tuned by changing the graphene-shell chemical potential, depends on dielectric core radius $R$ and permittivity, as well as $\epsilon$. 
These parameters are responsible for cloaking activation at other frequencies. 
Therefore, the results presented in Fig.~\ref{fig05}(a) are devoted specifically to $R = 0.5\ \mu\mathrm{m}$, and $\epsilon_d = 3.9$.

In Fig.~\ref{fig05}(b) we present the scattering efficiency spectrum $Q_{sca}$ for graphene coated dielectric nanowire as a function of core permittivity $\epsilon_d$, when immersed in the air ($\epsilon = 1$), while keeping fixed $R = 0.5\ \mu\mathrm{m}$, and $\mu = 0.5\ \mathrm{eV}$. 
Negative values of the permittivity $\epsilon_d$ can be realized via semiconductor doping, e.g. n-InGaAs. 
In the spectral range of interest, dispersion of that semiconductor $\epsilon_d (\omega)$ can be very well described by the Drude model with the plasmon wavelength located at $\lambda_p = 5.41\ \mu\mathrm{m}$ \cite{Caldwell15}. 
Minimum of the scattering efficiency and the peak of the resonance both shift to longer wavelengths with an increase of the core permittivity. 
For the core permittivity $\epsilon_d$ that is only slightly higher than $\epsilon$, the invisibility window significantly widens at the rate $d \omega_\mathrm{inv} / d \epsilon_d$, which is notably higher than $d \omega_\mathrm{res} / d \epsilon_d$, for the resonance peak window. 
For instance, the scattering minimum spans from $11.89\ \mu\mathrm{m}$ to $16.62\ \mu\mathrm{m}$ in the interval where $\epsilon_d$ changes from $1.5$ to $2.0$, whereas the variation of the resonant peak is produced from a wavelength of $26.09\ \mu\mathrm{m}$ to $28.53\ \mu\mathrm{m}$.
Moreover, when $|\epsilon_d| < \epsilon$, the conducting graphene sheet can excite a resonant peak but the invisibility window vanishes.
On the other hand, the maximum and the minimum of the curve virtually coincide for an epsilon-near-zero surrounding material.
We point out that a nonneglecting value of the imaginary part of $\epsilon$ imposes a correction of Eqs.~(\ref{eq14}) and (\ref{eq15}), thus preventing the peak-valley collapse. 
Inversely, high values of the environment permittivity enable to move the resonant peak further from the invisibility window, as depicted in Fig.~\ref{fig05}(b) for $\epsilon = 4$.

\begin{figure}[t]
	\centering
	\includegraphics[width=0.7\linewidth]{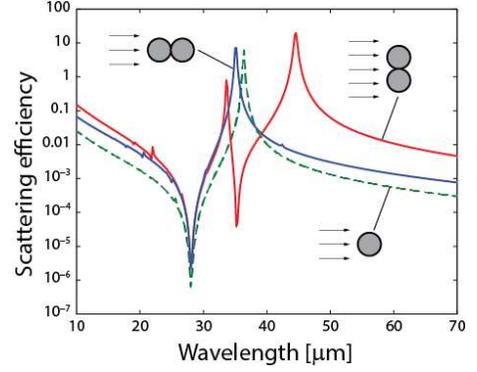}
	\caption{FEA-based numerical evaluation of the scattering efficiency of a cluster of two graphene-coated nanowires of radius $R = 0.5\ \mu\mathrm{m}$, core permittivity $\epsilon_d = 3.9$, and graphene chemical potential $\mu = 0.5\ \mathrm{eV}$, immersed in air.
		The nanowire cluster is oriented transversally (red solid line) and along (blue solid line) the propagation direction of the incident light.
		For comparison, we include the spectrum of $Q_\mathrm{sca}$ for a single nanoparticle in dashed green line.}
	\label{fig06}
\end{figure}

Next, we would like to analyze the scattering efficiency of the clusters of two invisible graphene-coated nanocylinders. 
In Fig.~\ref{fig06} we present the spectrum $Q_\mathrm{sca}$ that corresponds to dimmer nanowires, each of the radius $R = 0.5\ \mu\mathrm{m}$ and permittivity $\epsilon_d = 3.9$ over-coated by a uniform-graphene shell with chemical potential $\mu = 0.5\ \mathrm{eV}$. 
The pair of nanowires can be oriented either along the propagation direction of the incident plane wave, or perpendicularly. 
It is assumed that the axis of the cylinders is always in the $z$-direction. 
The scattering efficiency is numerically evaluated by using COMSOL Multiphysics, which is a commercially available solver of the Maxwell's equations based on the finite element analysis (FEA). 
In the first case, a set of a peak resonance and a scattering minimum, found at $\lambda_\mathrm{res} = 35.2\ \mu\mathrm{m}$ and $\lambda_\mathrm{inv} = 28.0\ \mu\mathrm{m}$ respectively, stands out in the scattering lineshape which is comparable to the spectral pattern for a single graphene-coated nanowire, except maybe in an increased efficiency magnitude.
For the case of two directly adjacent coated nanowires illuminated by light polarized along the interparticle $y$ axis, two minima are identified at wavelengths $\lambda_\mathrm{inv\ 1} = 28.0\ \mu\mathrm{m}$, coinciding with the previously analyzed invisibility window, and $\lambda_\mathrm{inv\ 2} = 35.2\ \mu\mathrm{m}$.
In addition, two resonant peaks are traced at $\lambda_\mathrm{res\ 1} = 33.6\ \mu\mathrm{m}$ and $\lambda_\mathrm{res\ 2} = 44.6\ \mu\mathrm{m}$.
The strong coupling between the graphene plasmons of each individual coated nanowire is well described by a hybridization picture relying on the analogy between plasmons and the wave functions of simple quantum systems \cite{Halas11}.

Finally, it is worth noting that scattering cancellation can be expected to appear at more different frequencies when dealing with clusters formed by a higher number of graphene-coated nanowires. 
Previously proposed alternate multi-frequency cloaking mechanism is based on several graphene layers implementation at different gate charges \cite{Farhat13}.

\section{Conclusions}

We have demonstrated that the scattering behavior of a sufficiently thin graphene-coated nano-cylinder reminiscences a solid nanowire with a bulk conductivity equal to the ratio of graphene surface conductivity and the wire radius. 
In that case, some analytical expressions (see equations (\ref{eq10})--(\ref{eq15})) have been derived enabling the characterization of the spectral window for the invisibility cloaking. 
We have examined the possibility of tuning the invisibility window by monitoring the applied gate voltage to control the chemical potential of graphene. 
Moreover, we reveal that the spectral distribution of scattering efficiency, for the high index ratio between the core and the surrounding medium, approaches the well-known Fano resonance line shape. 
We propose an adequate multi-frequency invisibility mechanism that is based on a hybrid picture with clusters of several invisible graphene-coated nano-cylinders. 
Our findings can be straightforwardly applied to more complex graphene coating structures \cite{Chen13b,Danaeifar16}. 
Finally, we would like to emphasize that the proposed scheme seems to be tunable enough to be implemented in ultra-thin reconfigurable cloaking devices. 
Within the THz range of frequencies this can be applied to low-noise sensing and imaging.


\end{document}